\documentclass[twocolumn,preprintnumbers,amsmath,amssymb,prl,floatfix]{revtex4}

\usepackage{graphicx}
\usepackage{dcolumn}
\usepackage{bm}
\usepackage{epsfig}
\usepackage{bm}
\newcommand{\beeq}{\begin{equation}}
\newcommand{\eneq}{\end{equation}}
\newcommand{\bear}{\begin{array}}
\newcommand{\enar}{\end{array}}
\newcommand{\bdi}{\begin{displaymath}}
\newcommand{\edi}{\end{displaymath}}

\begin{document}

\title{Inert states of spin-$S$ systems}

\author{H. M\"akel\"a and K.-A. Suominen}
\affiliation{Department of Physics, University of Turku, 
FI-20014 Turun yliopisto, Finland}


\begin{abstract}
We present a simple but efficient geometrical method for determining the inert states of spin-$S$ systems. It can be used if the system is described by a spin vector of a spin-$S$ particle and its energy is invariant in spin rotations and phase changes. Our method is applicable to an arbitrary $S$ and it is based on the representation of a pure spin state of a spin-$S$ particle in terms of $2S$ points on the surface of a sphere. We use this method to find candidates for some of the ground states of spinor Bose-Einstein condensates.
\end{abstract}

\maketitle

States which are stationary points of the energy regardless of the exact form of the energy functional are called inert states~\cite{Vollhardt90}. A stationary point may change, say, from a local maximum to a global minimum as the parameters characterizing the energy vary, but for all parameter values the inert state remains nevertheless a stationary point of energy. 
The independence of the details of the energy functional is related to the fact that inert states are determined by the symmetry group of the energy and the symmetry properties of the order-parameter alone. Inert states have been studied in the context of superfluid $^3$He, where the analytical minimization of the energy functional is a very complicated task, but the   
inert states can be calculated straightforwardly~\cite{Barton74,Vollhardt90}, see also \cite{Ozaki85}. It should be noted that sometimes the form of the ground state may depend on the parameters of the energy, in which case the ground state cannot be an inert state. It is therefore possible that not all stationary points of the energy are inert states, which is the case in $^3$He \cite{Vollhardt90,Bruder86}. For spin-$S$ systems in general, however, the calculation of inert states can also be tedious, as recently demonstrated for integer spin-values $S=1-4$~\cite{Yip06}, and the task seems to become increasingly complicated as the value of $S$ increases. In this paper we present a simple geometrical description that allows one to determine the inert states of spin-$S$ systems. The method is applicable regardless of the value of spin and by using it we explicitly construct numerous inert states. Almost all inert states for $S=1,2,3$ systems are ground states of the presently realized spinor Bose-Einstein condensates \cite{Yip06,Ho98,Ohmi98,Ciobanu00,Santos06,Diener06,Makela07}, which suggests that such states are also ground states of spinor condensates when $S$ is larger than three. 
   
Our way of calculating inert states is based on a geometrical representation of spin states.    
It is well known that a pure spin state of a spin-$1/2$ particle can be characterized by a point on the surface of the Bloch sphere. Moreover, a pure spin state $\xi$ of a spin-$S$ particle can be written in terms of $2S$ pure spin states of spin-$1/2$ particles \cite{Majorana32,Bloch45}. Combining these results, we see that $\xi$ can be expressed by $2S$ points on the surface of the Bloch sphere, which provides the geometrical description we are interested in. This decomposition was originally used to study the motion of a spin-$S$ particle in a magnetic field \cite{Majorana32,Bloch45}. Recently it has been used to calculate the symmetries of the ground states of spinor condensates with $S=2$ and $S=3$ \cite{Barnett06,Barnett07}, see also Refs. \cite{Barnett06b,Yip06b}. This decomposition is well suited for such purpose, since it shows explicitly how spin vectors change in rotations. Rotating $\xi$ corresponds to rotating the point configuration describing this vector with respect to the same axis and through the same angle. Let $\xi=\sum_{M=-S}^S \xi_M|S,M\rangle$ be a spin vector of a spin-$S$ particle. The points on the surface of the Bloch sphere characterizing $\xi$ can be obtained from the equation \cite{Majorana32,Bloch45}
\begin{eqnarray}\label{Eq1}
\sum_{M=-S}^{S}
   {2S \choose S+M}^{1/2}x^{S+M}\xi_{M}=A\prod_{k=1}^{2S}(\alpha_k x+\beta_k) =0.  
\end{eqnarray}
In the second expression we have chosen $A\in\mathbb{C}$ in such a way that we can define 
$\alpha_k=\cos(\theta_k/2) e^{-i\varphi_k/2}$ and $\beta_k=\sin(\theta_k/2) e^{i\varphi_k/2}$ for all $k$.
Every pair $(\alpha_k,\beta_k)$ determines a point on the Bloch sphere located at $(\theta_k,\varphi_k)$. In the rest of the paper we call the polynomial of Eq.~(\ref{Eq1}) the characteristic polynomial of $\xi$. If we are given two sets of points, then the characteristic polynomial of the configuration consisting of both sets 
is obtained by multiplying the characteristic polynomials of the two sets. Equation~(\ref{Eq1}) shows that the point distribution is the same for $\xi$ and $h\,\xi$, where $h$ is an arbitrary non-zero complex number. 

We now move on to calculating the inert states. We assume that the system is described by a spin vector of a spin-$S$ particle, denoted by $\xi$, and that the energy functional is  
invariant in rotations and phase transformations of $\xi$. Thus the symmetry group of the energy is $G=U(1)\times SO(3)$, where $SO(3)$ acts via its irreducible $2S+1$ dimensional representation and the action of $U(1)$ corresponds to multiplication by a complex number of modulus one. We define the isotropy group of $\xi$ by $H_\xi=\{g\in G\, |\, g\cdot\xi=\xi\}$, i.e., it consists of those elements of $U(1)\times SO(3)$ which leave $\xi$ fixed. Two isotropy groups, say $H_\xi$ and $H_{\xi'}$, are conjugate if $H_\xi=g H_{\xi'} g^{-1}=\{ghg^{-1}\,|\, h\in H_{\xi'}\}$ for some $g\in G$. In Ref.~\cite{Michel80} it has been shown that 
$\xi$ is an inert state if, for every $\xi'$ infinitesimally close to $\xi$, $H_\xi$ and $H_{\xi'}$ are not conjugate. Here it is assumed that $\xi'$ cannot be obtained from $\xi$ by a rotation and a phase change and that $||\xi||=||\xi'||$. In the present case each $\xi$ determines some point configuration on the surface of a sphere. A small variation in the positions of the points leads to a small change in $\xi$. Therefore, if all small changes necessarily lead to a new isotropy group which is not conjugate to $H_\xi$, then $\xi$ is an inert state. The elements of the isotropy group $H_\xi$ consists of pairs $(u,r)$, where $u\in U(1)$ and $r\in SO(3)$. The latter are the elements of some subgroup $K$ of $SO(3)$. $K$ can be determined by finding out all rotations which leave the point configuration invariant.  
We call $K$ the symmetry group or symmetry of the point configuration. It should not be mistaken for the isotropy group, see Ref.~\cite{Yip06b}. 

We find the inert states by determining point configurations which cannot be modified without changing the symmetry group $K$. However, even if $K$ is invariant as the positions of the points are varied, it is possible that the elements $u\in U(1)$, and therefore the isotropy group, change. Let $\mathbf{s}$ be some symmetry axis of the point configuration. Because of the rotational degree of freedom, we can choose $\mathbf{s}=\mathbf{z}$. Then $e^{-i\alpha_{rot} S_z}\xi = e^{i\delta}\xi$ for some $\alpha_{rot}$. Writing $\xi=\sum_{M=-S}^S\,\xi_{M}|S,M\rangle$ we get $e^{-i\alpha_{rot} M}\xi_M= e^{i\delta}\, \xi_{M}$ for all $M$. Assume that $p_n$ ($p_s$) of the points characterizing $\xi$ are at the north (south) pole. Then Eq.~(\ref{Eq1}) shows that the highest order term in the characteristic polynomial is $x^{2S-p_s}$ and the lowest order term is $x^{p_n}$. Therefore $\xi_{S-p_s},\xi_{-S+p_n}\not =0$ and we get  
\beeq\label{Rot}
\bear{ll}
\delta&=(S-p_n)\alpha_{rot}\ \,\,\,\, (\textrm{mod }2\pi)\\
&=(p_s-S)\alpha_{rot}\ \,\,\,\, (\textrm{mod }2\pi). 
\enar
\eneq
A similar result has been presented in Refs.~\cite{Barnett06b,Barnett07}. If $(u,r)\in H_\xi$ and the rotation angle of $r$ is $\alpha_{rot}$, then $u=e^{-i\delta}$. We see that if the elements of the $SO(3)$ part of $H_\xi$ remain unchanged as the points move, the $U(1)$ parts can change only if the number of points on a symmetry axis changes. If the smallest non-zero rotation angle leaving the points fixed is $\alpha_{rot}=2\pi/n$, then the only way to move points from the symmetry axis without affecting the symmetry group is to remove symmetrically $kn$ points from the axis. Here $k=1,2,\ldots$.  Equation~(\ref{Rot}) shows that in this case the $U(1)$ parts of $H_\xi$ remain unchanged, so it is possible to change the phase terms only if also the symmetry group changes. When calculating inert states we can thus concentrate only on the changes in the symmetry group. 

The possible symmetry groups are the continuous groups $O(2),SO(2)$, and the finite groups $I,C_n,D_n,T,O$ and $Y$. $I$ is the trivial group consisting of the identity element alone, 
$C_n$ ($D_n$) is the  cyclic (dihedral) group of order $n$, $T$ is the symmetry group of the tetrahedron, $O$ that of the octahedron or the cube, and $Y$ that of the icosahedron and the dodecahedron. We now calculate which of these subgroups determine inert states. We first  
find point configurations that have some given group as their symmetry group, and then see which of these cannot be modified without changing the symmetry group. In this context Ref.~\cite{Bacry74} is very useful. In the following we choose the $z$-axis to be one of the symmetry axes. 
 
{\it Orthogonal group $O(2)$.} The point configuration has to be invariant in arbitrary rotations about the $z$-axis and in a rotation through $\pi$ about an axis in the $xy$-plane. Therefore there has to be an equal number of points at the north and south poles and no points elsewhere. Clearly any change in the relative positions of the points will change the symmetry group, so this state is an inert state. Now $(\theta_k,\varphi_k)=(0,0)$ for $k=1,\ldots ,S$, $(\theta_k,\varphi_k)=(\pi,0)$ for $k=S+1,\ldots ,2S$, and the spin vector obtained from Eq.~(\ref{Eq1}) is $\xi^S_{O(2)}= |S,0\rangle$. This vector is clearly possible only if $S$ is an integer. Here and in the rest of the paper the subscript of $\xi$ gives the symmetry group 
of the point distribution and the superscript gives the value of the spin. 

{\it Special orthogonal group $SO(2)$.} This symmetry is present if there is only one symmetry axis and the point configuration is invariant in arbitrary rotations about this axis. Thus all the points have to be at the poles.  The number of points at different poles cannot be equal, 
because then the configuration would be invariant under $O(2)$. Also now the symmetry group changes if the relative positions of the points are changed, so this is an inert state. If there are $S+M$ points at the north pole, then  $(\theta_k,\varphi_k)=(0,0)$ for $k=1,\ldots ,S+M$ and $(\theta_k,\varphi_k)=(\pi,0)$ for $k=S+M+1,\ldots, 2S$. These give $\xi^S_{SO(2)}=|S,M \rangle$, where $M>0$. It is sufficient to consider positive $M$ only, since $|S,M\rangle$ and $|S,-M\rangle$ can be obtained from each other by a rotation. Now inert states can exist also when $S$ is a half integer. 

\begin{figure}[htb]
\includegraphics[scale=.85]{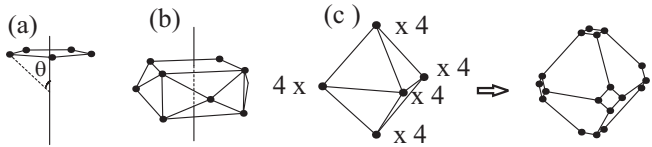}   
\caption{\label{Noninert} Some non-inert states. In (a) $\theta$ and in (b) the height of the box can be changed without affecting the symmetry. In (c) we show the truncation of an octahedron.} 
\end{figure}

{\it Cyclic group $C_n$.} The only symmetry axis is the $z$-axis, and the system is invariant in rotations through $2\pi k/n$ about this axis. There has to be a point at $(\theta,\varphi)=(\theta_C,\varphi_C+2\pi k/n)$ for some fixed $\theta_C\in (0,\pi)$, $\varphi_C\in [0,2\pi/n)$, and for each $k=1,\ldots ,n$. There may be points also elsewhere, as long as the configuration is invariant under $C_n$. The value of $\theta_C$ can always be changed by some $\delta\theta_C$  without altering the symmetry group. This change cannot be achieved by a rotation. Therefore the cyclic groups do not determine any inert states. In Fig.~\ref{Noninert}(a) we show an example of the $C_5$ symmetry in an $S=5/2$ system. 

{\it Dihedral group $D_n$.} Because $C_n$ is a subgroup of $D_n$, there has to be a point at 
$(\theta,\varphi)=(\theta_D,\varphi_D+2\pi k/n)$.  
Now there are $n$ symmetry axes in the $xy$-plane, so there is an equal number of points at the north and the south pole.
If there are points outside the $xy$-plane and the poles, it is possible to change the $\theta$-coordinate of these points without changing the symmetry group, so these configurations do not give inert states. In Fig.~\ref{Noninert}(b) this is illustrated for a configuration with $D_4$ symmetry. The case where there are points only in the $xy$-plane and at the poles can be divided into three cases. Below $k=1,2,\ldots ,n$, $l=1,2,\ldots$, and $\varphi_l\in[0,2\pi/n)$ (I.) $n$ or more points at each pole and $ln$ points in the $xy$-plane. Now $n$ points can be moved from each pole towards the equatorial plane without changing the symmetry, see Fig.~\ref{Noninert}(a). (II.) Less than $n$ points at each pole, two or more points at $(\theta,\varphi)=(\pi/2,\varphi_1+2\pi k/n)$, and $ln$ points elsewhere in the $xy$-plane.  It is possible to move one point up and one down by the same amount at each $(\pi/2,\varphi_1+2\pi k/n)$ without changing the symmetry group. (III.) Less than $n$ points at each pole and one point at each $(\theta,\varphi)=(\pi/2,\varphi_l+2\pi k/n)$, $\varphi_l\not =\varphi_{l'}$ if $l\not = l'$.  When the symmetry axes in the $xy$-plane are taken into attention, it turns out that $l=1$ is the only possibility~\cite{Makela}. 

We therefore conclude that an inert state with $D_n$ symmetry is possible only if there is one point at $(\theta,\varphi)=(\pi/2,2\pi k/n)$ for $k=1,2,\ldots, n $ and $p<n$ points at each pole (we have chosen $\varphi_1=0$).  It follows that in a spin-$S$ system there is an inert state corresponding to $D_n$ only if $(2S+2)/3 \leq n\leq 2S$. Furthermore, the condition $n=2S-2l$, where $l$ is a non-negative integer, has to hold. This is because, for fixed $S$, $n$ can be decreased only by putting an equal number of points at each pole. Therefore for integer (half integer) spin only even (odd) values of $n$ are possible, the smallest value of $n$ being given by the smallest even (odd) integer bounding $(2S+2)/3$ above. It is to be noted that if $S=3$, $n=4$, the symmetry of the point configuration is not $D_4$, but that of an octahedron. If $S=2$, $n=2$ or $n=4$, the symmetry is $D_4$. The $n=4$ case corresponds to a square in the horizontal plane, while $n=2$ gives a square in a vertical plane. We define $\xi^S_{D_{2S-2l}}=|S,S-l\rangle+|S,-S+l\rangle$. Using Eq.~(\ref{Eq1}) it is easy to verify that this is a vector with $D_{2S-2l}$ symmetry. Dihedral inert states exist if $S>1$ for both integer and half integer $S$.  

\begin{figure}[htb]
\includegraphics[scale=.85]{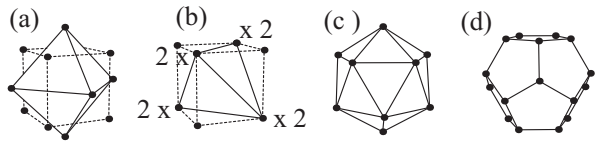}
\caption{\label{Inert}  Examples of inert states. 
In (a) cube and octahedron and in (b) cube and tetrahedron inscribed in a sphere simultaneously. In (a) the symmetry is that of octahedron/cube and in (b) that of tetrahedron. In (c) icosahedron and in (d) dodecahedron.} 
\end{figure}

The remaining possible symmetry groups are tetrahedral, octahedral and icosahedral groups. 
We calculate only those inert states which are obtained by placing points at the vertices of the convex regular polyhedra. The latter are often called the Platonic solids, and they are the tetrahedron, octahedron, cube, icosahedron and dodecahedron. Any change in the relative positions of the vertices changes the symmetry group (assuming that rotations are excluded).     
It is possible that there are also other ways to distribute points on the surface of a sphere with, say, octahedral symmetry than to place them at the vertices of the octahedron or cube. However, these distributions correspond to polyhedra which are not regular, and therefore the points can probably be moved without changing the symmetry group. Platonic solids have an even number of vertices, which means that inert states 
with half integer $S$ are not possible.   

{\it Octahedral group $O$.}  Octahedron and cube can be inscribed in a sphere simultaneously so that the combined point configuration has octahedral symmetry, see Fig.~\ref{Inert}(a). This is obtained by putting the vertices of the cube at $(\theta,\varphi)=(\theta_{Cube},\pi k/2)$, $(\pi-\theta_{Cube},\pi k/2)$, and those of the octahedron at $(\theta,\varphi)=(0,0),(\pi,0),(\pi /2,\pi (1+2k)/4)$. Here $k=1,2,3,4$ and 
$\theta_{Cube}=\tan^{-1}(\sqrt{2})$. The corresponding characteristic polynomials are   
 $x^8-14 x^4+1$ and $x^5 +x$ for the cube and octahedron, respectively. The vectors obtained from these polynomials are $\xi_{Cube}^4=\sqrt{5}|4,4\rangle -\sqrt{14}|4,0\rangle +\sqrt{5} |4,-4\rangle$ and $\xi^3_{Octa}=|3,2\rangle+|3,-2\rangle$. If there are $m$ ($n$) points at each vertex of a cube (octahedron), the characteristic polynomial becomes  $(x^8-14 x^4+1)^m (x^5+x)^n$.  Corresponding spin vector can be calculated by comparing this polynomial with the first expression in Eq.~(\ref{Eq1}). For example, if $m=n=1$, the characteristic polynomial becomes $x^{13}-13 x^9 -13 x^5+x$, which gives $\xi^7_{Cube+Octa}=\sqrt{11}|7,6\rangle-\sqrt{13}|7,2\rangle -\sqrt{13}|7,-2\rangle+\sqrt{11}|7,-6\rangle$. It is important to notice that  
if there are four or more points at each vertex of the octahedron, it is possible to 
truncate it continuously, see Fig.~\ref{Noninert}(c). Truncation does not change the symmetry or isotropy group but it changes the state, so the octahedron determines an inert state only if there are less than four points at each vertex. Similarly the cube can be truncated if there are three or more points at each vertex. Combining these results, we see that inert states determined by the cube and octahedron are possible if $S=3m+4n$, where $m+n\geq 1$, $m\leq 3$, and $n\leq 2$.  

{\it Tetrahedral group $T$.} Tetrahedron has four vertices, which we choose to be at $(\theta,\varphi)=(\theta_{Tetra},\pi/4),(\theta_{Tetra},5 \pi/4),(\pi-\theta_{Tetra},3\pi/4),(\pi-\theta_{Tetra},7\pi/4)$, where $\theta_{Tetra}=\tan^{-1}(\sqrt{2})$. The characteristic polynomial is $x^4+2i \sqrt{3} x^2+1$, which gives $\xi^2_T=|2,2\rangle +i\sqrt{2}|2,0\rangle +|2,-2\rangle$. The tetrahedron, octahedron, and cube can be inscribed in a sphere simultaneously, see Figs.~\ref{Inert}(a) and \ref{Inert}(b). The vertices can be chosen to be at the points given above. The resulting configuration  has tetrahedral symmetry, which therefore can be present at least if $S=2l+3m+4n$. Here $l\geq 1$ and $m,n\geq 0$. 
Because of the restrictions imposed by truncation, the conditions $l+n\leq 2$ and $m\leq 3$ have to hold. 

{\it Icosahedral group $Y$.} Icosahedron can be truncated if there are five or more points at each vertex, while for dodecahedron this is possible if there are more than two points at each vertex, see Figs.~\ref{Inert}(c) and \ref{Inert}(d). Thus these symmetries are possible if $S=6m+10n$, where $m+n\geq 1$, $m\leq 4$, and $n\leq 2$.  We obtain  $\xi^6_{Ico}=\sqrt{7}|6,5\rangle-\sqrt{11} |6,0\rangle -\sqrt{7}|6,-5\rangle$ and $\xi^{10}_{Dode}=\sqrt{17}|10,10\rangle +\sqrt{57}|10,5\rangle +\sqrt{\frac{247}{11}}|10,0\rangle -\sqrt{57}|10,-5\rangle+\sqrt{17} |10,-10\rangle$, details of the calculation will be presented in Ref.~\cite{Makela}. 

\begin{table}
\caption{The inert states for $S=1-4$. The only symmetry group that is not manifested here 
is the icosahedral group.\label{Tab}}
\begin{ruledtabular}
\begin{tabular}{|c|c|}
Spin & Inert states\\
\hline
$S=1$ & $\xi_{SO(2)}=|1,1\rangle,\ \xi_{O(2)}=|1,0\rangle$\\
\hline
$S=2$ & $\xi_{SO(2)}=|2,2\rangle$ and $|2,1\rangle,$ $\xi_{D_4}=|2,2\rangle+|2,-2\rangle$,\\
&$\xi_{O(2)}=|2,0\rangle$, and $\xi_{Tetra}=|2,2\rangle+i\sqrt{2}|2,0\rangle+|2,-2\rangle$ \\
\hline
$S=3$ & $\xi_{SO(2)}=|3,3\rangle, |3,2\rangle,$ and $|3,1\rangle$, $\xi_{O(2)}=|3,0\rangle,$\\
& $\xi_{D_6}=|3,3\rangle+|3,-3\rangle,$ and $\xi_{Octa}=|3,2\rangle+|3,-2\rangle$\\
\hline
$S=4$ & $\xi_{SO(2)}=|4,4\rangle,|4,3\rangle,|4,2\rangle,$ and $|4,1\rangle$, $\xi_{O(2)}=|4,0\rangle$,\\
& $\xi_{D_8}=|4,4\rangle +|4,-4\rangle, \xi_{D_6}=|4,3\rangle+|4,-3\rangle$\\
& $\xi_{D_4}=|4,2\rangle +|4,-2\rangle$, $\xi_{Tetra}=\sqrt{7}|4,4\rangle +2i\sqrt{3}|4,2\rangle$\\
& $-\sqrt{10}|4,0\rangle +2i\sqrt{3}|4,-2\rangle+\sqrt{7}|4,-4\rangle$.\\
& $\xi_{Cube}=\sqrt{5}|4,4\rangle-\sqrt{14}|4,0\rangle +\sqrt{5}|4,-4\rangle$\\
\end{tabular}
\end{ruledtabular}
\end{table}

In Table~\ref{Tab} we list the inert states for integer-valued spin $S=1-4$. All except $\xi_{D_4}=|4,2\rangle+|4,-2\rangle$ have also been obtained using a different method~\cite{Yip06}. This state was omitted in Ref.~\cite{Yip06} because it was assumed that the isotropy group of $\xi_{D_4}$ is a subgroup of the isotropy group of $\xi_{D_8}$~\cite{Note}. But we find that this is not the case, as can be seen by considering a rotation through $\pi/2$ about the $z$-axis. In this case Eq.~(\ref{Rot}) shows that for $D_4$ symmetry $\delta=\pi$, while for $D_8$ symmetry $\delta=0$.

It is interesting to compare the inert states of Table~\ref{Tab} with the ground states of spinor Bose-Einstein condensates with $S=1,2,3$. For $S=1$ there are no other ground states than the inert states~\cite{Ho98,Ohmi98}. If $S=2$ it seemed initially that also non-inert ground states are possible due to an accidental degeneracy~\cite{Ciobanu00,Ueda02}, but recently it has been shown that due to quantum and thermal fluctuations this degeneracy is lifted, 
after which only inert ground states can occur~\cite{Song07,Turner07}. However, there is one inert state, namely $|2,1\rangle$, which is never a ground state. If $S=3$, the inert states are ground states but also numerous other ground states are possible~\cite{Diener06,Santos06,Makela07}. These results suggest that the inert states for $S>3$ are ground states of spinor condensates at least for some values of the scattering lengths. This has interesting consequences considering topological defects, as it shows that monopoles made possible by the $|S,0\rangle$ ground state and non-Abelian vortices arising from the $D_n$-symmetric states are possible in spinor condensates regardless of the value of spin.   
In this paper we have not considered the time-reversal symmetry. Its effects will be discussed elsewhere \cite{Makela}. To conclude, we have presented a simple geometrical method to calculate the inert states of systems which are described by a single particle wave function of a spin-$S$ particle and which have $U(1)\times SO(3)$ invariant energy functional. The possible inert states can be identified and classified by their symmetry groups. 

The authors acknowledge the financial support of the Academy of Finland (Grant No.~115682). H. M. was supported by the Finnish Academy of Science and Letters through the Vilho, Yrj\"o and Kalle V\"ais\"al\"a Foundation.


\begin{thebibliography}{11}

\bibitem{Vollhardt90} {\it The Superfluid Phases of Helium 3}, D. Vollhardt and P. W\"olfle, 
Taylor and Francis, London, 1990. 

\bibitem{Barton74} G. Barton and M. A. Moore, J. Phys. C {\bf 7}, 4220 (1974); {\it ibid.} {\bf 8}, 970 (1975).

\bibitem{Ozaki85} M. Ozaki, K. Machida, and T. Ohmi, Prog. Theor. Phys. {\bf 74}, 221 (1985); {\it ibid.} {\bf 75}, 442 (1986).

\bibitem{Bruder86} C. Bruder and D. Vollhardt, Phys. Rev. B {\bf 34}, 131 (1986). 

\bibitem{Yip06} S.-K. Yip, Phys. Rev. A {\bf 75}, 023625 (2007). 

\bibitem{Ho98}T.-L. Ho, Phys. Rev. Lett. {\bf 81}, 742 (1998).

\bibitem{Ohmi98} T. Ohmi and K. Machida, J. Phys. Soc. Jpn. {\bf 67}, 1822 (1998). 

\bibitem{Ciobanu00} C. V. Ciobanu, S.-K. Yip, and T.-L. Ho, Phys. Rev. A {\bf 61}, 033607 (2000). 

\bibitem{Diener06} R. B. Diener and T.-L. Ho, Phys. Rev. Lett. {\bf 96}, 190405 (2006).

\bibitem{Santos06} L. Santos and T. Pfau, Phys. Rev. Lett. {\bf 96}, 190404 (2006).

\bibitem{Makela07} H. M\"akel\"a and K.-A. Suominen, Phys. Rev. A {\bf 75}, 033610 (2007). 

\bibitem{Majorana32} E. Majorana, Nuovo Cimento {\bf 9}, 43 (1932). 

\bibitem{Bloch45} F. Bloch and I. I. Rabi, Rev. Mod. Phys. {\bf 17}, 237 (1945). 

\bibitem{Barnett06} R. Barnett, A. Turner, and E. Demler, Phys. Rev. Lett. {\bf 97}, 180412 (2006). 

\bibitem{Barnett07} R. Barnett, A. Turner, and E. Demler, Phys. Rev. A {\bf 76}, 013605 (2007).

\bibitem{Barnett06b} R. Barnett, A. Turner, and E. Demler, cond-mat/0612099.

\bibitem{Yip06b} S.-K. Yip, cond-mat/0611426. 

\bibitem{Michel80}L. Michel, C. R. Acad. Sci., Ser. A {\bf 272}, 433 (1971); 
Rev. Mod. Phys. {\bf 52}, 617 (1980). 

\bibitem{Bacry74} H. Bacry, J. Math. Phys. {\bf 15}, 1686 (1974). 

\bibitem{Makela} H. M\"akel\"a and K.-A. Suominen, in preparation.

\bibitem{Note} In Ref. \cite{Yip06} it was argued that in this case $\xi_{D_4}$ cannot be an inert state. 

\bibitem{Ueda02} M. Ueda and M. Koashi, Phys. Rev. A {\bf 65}, 063602 (2002). 

 \bibitem{Song07} J. L. Song, G. W. Semenoff, and F. Zhou, Phys. Rev. Lett. {\bf 98}, 160408 (2007). 

\bibitem{Turner07} A. M. Turner {\it et al.},  
Phys. Rev. Lett. {\bf 98}, 190404 (2007). 

\end{thebibliography}
\end{document}